\documentclass[sigconf]{acmart}
\renewcommand\footnotetextcopyrightpermission[1]{} % removes footnote with conference information in first column
\usepackage{booktabs} 
\usepackage{todonotes}
\usepackage{balance}

\begin{document}
\title{Competitive Video Retrieval with vitrivr at the Video Browser Showdown 2018 -- Final Notes}

\author{Luca Rossetto, Ivan Giangreco, Ralph Gasser and Heiko Schuldt}
\affiliation{%
  \institution{Department of Mathematics and Computer Science\\University of Basel}
  \city{Basel, Switzerland}
}
\email{{firstnale.lastname}@unibas.ch}

\begin{abstract}
This paper presents an after-the-fact summary of the participation of the vitrivr system~\cite{rossetto2016vitrivr} to the 2018 Video Browser Showdown~\cite{cobarzan2017interactive}. A particular focus is on additions made since the original publication~\cite{rossetto2018competitive} and the systems performance during the competition.
\end{abstract}


\maketitle

\section{System Overview}

vitrivr is an open-source content-based multimedia retrieval stack, capable of retrieving not only video, but also images, audio, and 3D models~\cite{rossetto2018open}. The vitrivr stack is the open-source continuation of the IMOTION system, which has participated in the Video Browser Showdown for several years~\cite{rossetto2015imotion,rossetto2016imotion,rossetto2017enhanced}. vitrivr supports many different query modes such as Query-by-Sketch and Query-by-Example as well as text-based queries for text-on-screen, dialog, or semantic concepts. 
 
\begin{figure*}
\centering
\includegraphics[width=0.9\textwidth]{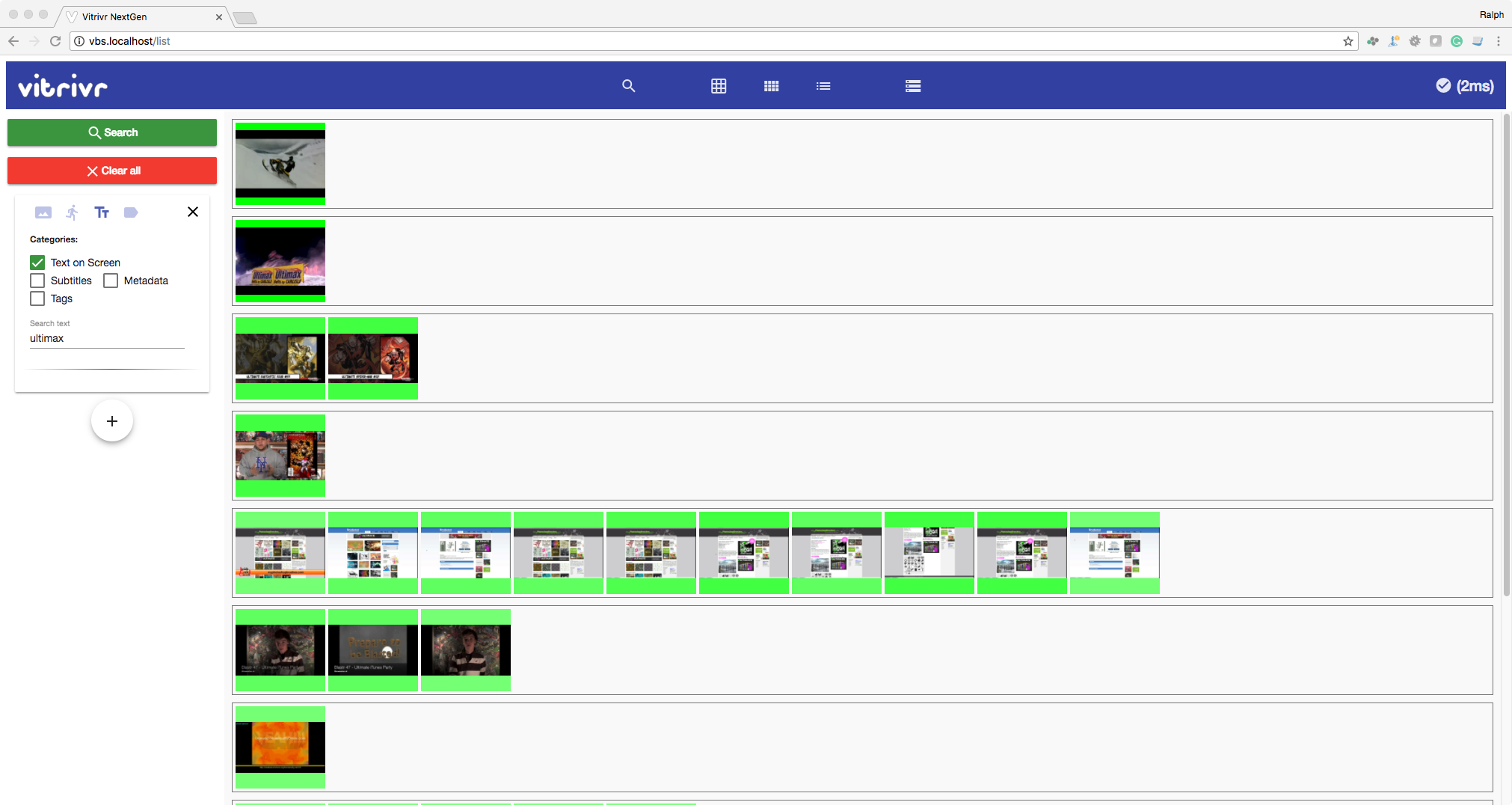}
\caption{Screenshot of the vitrivr user interface}
\label{fig:screenshot}
\end{figure*}

The mechanisms used for visual content-based queries are already detailed in previous publications~\cite{rossetto2014cineast,rossetto2015imotion,rossetto2016imotion,rossetto2017enhanced,rossetto2018competitive} and have remained the same also for the version used at the Video Browser Showdown 2018. Figure~\ref{fig:screenshot} shows a screenshot of the vitrivr user interface.

\section{Additions}
This section provides a brief overview of the additions that were made after the submission of the original publication~\cite{rossetto2018competitive}. They concern both the user interface and the textual features used.

In order to improve the usability and the retrieval effectiveness in a time-sensitive and competitive environment, the user interface was adapted. Since the user interface was originally designed to support retrieval across different types of multimedia, it was not optimized to display a high density of video data. Two additional data views have therefore been added to the interface to improve its capabilities in this regard. The first one fills the screen with representative thumbnail images of video segments and arranges them by their similarity to the query, independently of their video of origin. The second view, an example of which can be seen in Figure~\ref{fig:screenshot}, groups the same thumbnails by video and displays them in temporal order within each group. In these views, the result set can be expanded by manually loading neighboring segments of a given segment in order to gain a better overview of the video's content. Loading all segments of a video in such a way is also possible. Segments can also be annotated with different colors in order to quickly find them again when switching between views or re-ordering the segments based on different similarity criteria. The video player was extended to support higher playback speeds as well as submitting a result directly based on the current playback position.

With respect to the data used for retrieval, in particular to execute textual queries, two data sources were used: The spoken words were already provided with the collection as the result of an automated speaker recognition process. This ASR data was transformed into approximate subtitles by thresholding the probabilities associated with the recognized words. For object labeling and scene text recognition, the Google Cloud Vision API\footnote{\url{https://cloud.google.com/vision/}} was used. The text-based retrieval was performed using the Apache Solr\footnote{\url{http://lucene.apache.org/solr/}} text search platform with fuzzy queries in order to compensate for minor mis-detections and typos.

\section{System use and performance}
During the competition, both `expert' and `novice' users relied heavily on the text-based retrieval capabilities, at least in a first attempt. This was probably due to the little amount of time needed to specify a textual query as opposed to creating a visual sketch, which was rather time consuming. Text-based queries have also shown to work very well in cases where there is distinctive dialog in the target sequence or clearly readable text  visible on-screen. Textual queries also perform well when a detectable object is being displayed in the scene which is uncommon enough to serve as an effective filter while being sufficiently common to still be recognized by one of the available object detectors. 

In case none of these prerequisites for an effective textual query were present, the visual query served as a fallback option. During the competition, many tasks would not lend themselves well to text-based queries and for some of them, even visual queries proved largely ineffective. Some possible reasons for this are elaborated on in Section~\ref{sec:lessons-learned}. The overall placements for the different task types were as follows:

\begin{itemize}
\item 1$^{st}$ place in pre-competition textual KIS tasks (100/100 points). These points were not counted towards the final points as this was a separate session the day prior to the actual competition.
\item 2$^{nd}$ place in competition textual KIS tasks (55/100 points)
\item 4$^{th}$ place in visual expert KIS tasks (79/100 points)
\item 5$^{th}$ place in expert AVS tasks (64/100 points)
\item 7$^{th}$ place in novice AVS tasks (35/100 points)
\item no placement in novice visual KIS tasks (0/100 points)
\item 7$^{th}$ place overall (47/100)
\end{itemize}

\section{Lessons Learned}
\label{sec:lessons-learned}
The experiences during the competition as well as the subsequent analysis of the results uncovered several lessons for future participations, which can be summarized as follows:

\begin{itemize}
\item As in 2017, we used the master shot references provided with the IACC.3. While this worked well in 2017, there were problems with this segmentation in 2018. In many instances, the segmentation was too coarse so that the actual target scene was part of a much longer segment and could therefore not be found. 
\item The fuzzy search method used for textual queries, while adequate for ASR and OCR data, proved disadvantageous for concepts. For example, results for the query `toast' would include beach-settings as they were tagged with `coast'.
\item The actual browsing capabilities of the system turned out to be insufficient in situations where no sufficiently selective query could be formulated. In case of many possibly relevant results, the submission mechanism for AVS turned out to also be not very effective.
\item For several queries, the same elements dominated the result set independent of query variations. Methods to increase result diversity should be explored in the future.
\end{itemize}

\section{Conclusion}
The results show that the overall approaches used for the vitrivr stack are sound but that there is still room for improvement. The wide range in placement shows that the system does not yet reliably produce high quality results in a competitive setting. The learned lessons discussed above will serve as a guide for improving this reliability in the future.

\section*{Acknowledgements}
This work was partly supported by the Chist-Era project
IMOTION with contributions from the Swiss National Science Foundation (SNSF,
contract no. 20CH21\_151571). The authors would like to thank our `novice' user for operating the system during the novice tracks of the competition.

\balance 

\bibliographystyle{ACM-Reference-Format}
\bibliography{bibliography}


\begin{thebibliography}{8}

%%% the Web address.

\ifx \showCODEN    \undefined \def \showCODEN     #1{\unskip}     \fi
\ifx \showDOI      \undefined \def \showDOI       #1{#1}\fi
\ifx \showISBNx    \undefined \def \showISBNx     #1{\unskip}     \fi
\ifx \showISBNxiii \undefined \def \showISBNxiii  #1{\unskip}     \fi
\ifx \showISSN     \undefined \def \showISSN      #1{\unskip}     \fi
\ifx \showLCCN     \undefined \def \showLCCN      #1{\unskip}     \fi
\ifx \shownote     \undefined \def \shownote      #1{#1}          \fi
\ifx \showarticletitle \undefined \def \showarticletitle #1{#1}   \fi
\ifx \showURL      \undefined \def \showURL       {\relax}        \fi
% invisible to TeX
\providecommand\bibfield[2]{#2}
\providecommand\bibinfo[2]{#2}
\providecommand\natexlab[1]{#1}
\providecommand\showeprint[2][]{arXiv:#2}

\bibitem[\protect\citeauthoryear{Cob{\^a}rzan, Schoeffmann, Bailer, H{\"u}rst,
  Bla{\v{z}}ek, Loko{\v{c}}, Vrochidis, Barthel, and Rossetto}{Cob{\^a}rzan
  et~al\mbox{.}}{2017}]%
        {cobarzan2017interactive}
\bibfield{author}{\bibinfo{person}{Claudiu Cob{\^a}rzan},
  \bibinfo{person}{Klaus Schoeffmann}, \bibinfo{person}{Werner Bailer},
  \bibinfo{person}{Wolfgang H{\"u}rst}, \bibinfo{person}{Adam Bla{\v{z}}ek},
  \bibinfo{person}{Jakub Loko{\v{c}}}, \bibinfo{person}{Stefanos Vrochidis},
  \bibinfo{person}{Kai~Uwe Barthel}, {and} \bibinfo{person}{Luca Rossetto}.}
  \bibinfo{year}{2017}\natexlab{}.
\newblock \showarticletitle{Interactive video search tools: a detailed analysis
  of the video browser showdown 2015}.
\newblock \bibinfo{journal}{\emph{Multimedia Tools and Applications}}
  \bibinfo{volume}{76}, \bibinfo{number}{4} (\bibinfo{year}{2017}),
  \bibinfo{pages}{5539--5571}.
\newblock


\bibitem[\protect\citeauthoryear{Rossetto, Giangreco, Gasser, and
  Schuldt}{Rossetto et~al\mbox{.}}{2018a}]%
        {rossetto2018competitive}
\bibfield{author}{\bibinfo{person}{Luca Rossetto}, \bibinfo{person}{Ivan
  Giangreco}, \bibinfo{person}{Ralph Gasser}, {and} \bibinfo{person}{Heiko
  Schuldt}.} \bibinfo{year}{2018}\natexlab{a}.
\newblock \showarticletitle{Competitive Video Retrieval with vitrivr}. In
  \bibinfo{booktitle}{\emph{International Conference on Multimedia Modeling}}.
  Springer, \bibinfo{pages}{403--406}.
\newblock


\bibitem[\protect\citeauthoryear{Rossetto, Giangreco, Gasser, and
  Schuldt}{Rossetto et~al\mbox{.}}{2018b}]%
        {rossetto2018open}
\bibfield{author}{\bibinfo{person}{Luca Rossetto}, \bibinfo{person}{Ivan
  Giangreco}, \bibinfo{person}{Ralph Gasser}, {and} \bibinfo{person}{Heiko
  Schuldt}.} \bibinfo{year}{2018}\natexlab{b}.
\newblock \showarticletitle{Open-source column: content-based multimedia
  retrieval using vitrivr}.
\newblock \bibinfo{journal}{\emph{ACM SIGMultimedia Records}}
  \bibinfo{volume}{9}, \bibinfo{number}{3} (\bibinfo{year}{2018}),
  \bibinfo{pages}{8}.
\newblock


\bibitem[\protect\citeauthoryear{Rossetto, Giangreco, Heller, T{\u{a}}nase,
  Schuldt, Dupont, Seddati, Sezgin, Alt{\i}ok, and Sahillio{\u{g}}lu}{Rossetto
  et~al\mbox{.}}{2016a}]%
        {rossetto2016imotion}
\bibfield{author}{\bibinfo{person}{Luca Rossetto}, \bibinfo{person}{Ivan
  Giangreco}, \bibinfo{person}{Silvan Heller}, \bibinfo{person}{Claudiu
  T{\u{a}}nase}, \bibinfo{person}{Heiko Schuldt}, \bibinfo{person}{St{\'e}phane
  Dupont}, \bibinfo{person}{Omar Seddati}, \bibinfo{person}{Metin Sezgin},
  \bibinfo{person}{Ozan~Can Alt{\i}ok}, {and} \bibinfo{person}{Yusuf
  Sahillio{\u{g}}lu}.} \bibinfo{year}{2016}\natexlab{a}.
\newblock \showarticletitle{IMOTION--searching for video sequences using
  multi-shot sketch queries}. In \bibinfo{booktitle}{\emph{International
  Conference on Multimedia Modeling}}. Springer, \bibinfo{pages}{377--382}.
\newblock


\bibitem[\protect\citeauthoryear{Rossetto, Giangreco, and Schuldt}{Rossetto
  et~al\mbox{.}}{2014}]%
        {rossetto2014cineast}
\bibfield{author}{\bibinfo{person}{Luca Rossetto}, \bibinfo{person}{Ivan
  Giangreco}, {and} \bibinfo{person}{Heiko Schuldt}.}
  \bibinfo{year}{2014}\natexlab{}.
\newblock \showarticletitle{Cineast: a multi-feature sketch-based video
  retrieval engine}. In \bibinfo{booktitle}{\emph{Multimedia (ISM), 2014 IEEE
  International Symposium on}}. IEEE, \bibinfo{pages}{18--23}.
\newblock


\bibitem[\protect\citeauthoryear{Rossetto, Giangreco, Schuldt, Dupont, Seddati,
  Sezgin, and Sahillio{\u{g}}lu}{Rossetto et~al\mbox{.}}{2015}]%
        {rossetto2015imotion}
\bibfield{author}{\bibinfo{person}{Luca Rossetto}, \bibinfo{person}{Ivan
  Giangreco}, \bibinfo{person}{Heiko Schuldt}, \bibinfo{person}{St{\'e}phane
  Dupont}, \bibinfo{person}{Omar Seddati}, \bibinfo{person}{Metin Sezgin},
  {and} \bibinfo{person}{Yusuf Sahillio{\u{g}}lu}.}
  \bibinfo{year}{2015}\natexlab{}.
\newblock \showarticletitle{IMOTION—a content-based video retrieval engine}.
  In \bibinfo{booktitle}{\emph{International Conference on Multimedia
  Modeling}}. Springer, \bibinfo{pages}{255--260}.
\newblock


\bibitem[\protect\citeauthoryear{Rossetto, Giangreco, Tanase, and
  Schuldt}{Rossetto et~al\mbox{.}}{2016b}]%
        {rossetto2016vitrivr}
\bibfield{author}{\bibinfo{person}{Luca Rossetto}, \bibinfo{person}{Ivan
  Giangreco}, \bibinfo{person}{Claudiu Tanase}, {and} \bibinfo{person}{Heiko
  Schuldt}.} \bibinfo{year}{2016}\natexlab{b}.
\newblock \showarticletitle{vitrivr: A Flexible Retrieval Stack Supporting
  Multiple Query Modes for Searching in Multimedia Collections}. In
  \bibinfo{booktitle}{\emph{Proceedings of the 2016 ACM on Multimedia
  Conference}}. ACM, \bibinfo{pages}{1183--1186}.
\newblock


\bibitem[\protect\citeauthoryear{Rossetto, Giangreco, T{\u{a}}nase, Schuldt,
  Dupont, and Seddati}{Rossetto et~al\mbox{.}}{2017}]%
        {rossetto2017enhanced}
\bibfield{author}{\bibinfo{person}{Luca Rossetto}, \bibinfo{person}{Ivan
  Giangreco}, \bibinfo{person}{Claudiu T{\u{a}}nase}, \bibinfo{person}{Heiko
  Schuldt}, \bibinfo{person}{St{\'e}phane Dupont}, {and} \bibinfo{person}{Omar
  Seddati}.} \bibinfo{year}{2017}\natexlab{}.
\newblock \showarticletitle{Enhanced retrieval and browsing in the IMOTION
  system}. In \bibinfo{booktitle}{\emph{International Conference on Multimedia
  Modeling}}. Springer, \bibinfo{pages}{469--474}.
\newblock


\end{thebibliography}

\end{document}